\documentclass[journal]{IEEEtran}
\ifCLASSINFOpdf

\else

\fi

\usepackage{mathrsfs}
\usepackage{amsmath}
\usepackage{amsfonts}
\usepackage{amsthm}
\usepackage{amssymb}
\usepackage{graphicx}
\usepackage{indentfirst}
\usepackage{array}
\usepackage{cite}
\usepackage[linesnumbered,ruled,vlined]{algorithm2e}
\usepackage{multirow}
\usepackage{latexsym, bm}
\usepackage{booktabs}
\usepackage{color,xcolor}
\usepackage{epstopdf}

\newtheorem{lemma}{\bf{Lemma}}

\begin{document}

\title{\LARGE{Polar-Precoding: A Unitary Finite-Feedback Transmit Precoder for Polar-Coded MIMO Systems}}
\author{Jinnan Piao, \emph{Student Member, IEEE}, Kai Niu, \emph{Member, IEEE}, Jincheng Dai, \emph{Member, IEEE}, and Lajos Hanzo, \emph{Fellow, IEEE}
\thanks{Copyright (c) 2015 IEEE. Personal use of this material is permitted. However, permission to use this material for any other purposes must be obtained from the IEEE by sending a request to pubs-permissions@ieee.org.}
\thanks{This work is supported by National Key R\&D Program of China (No. 2018YFE0205501), the National Natural Science Foundation of China (No. 62071058 \& No. 62001049), China Post-Doctoral Science Foundation (No. 2019M660032), the
Engineering and Physical Sciences Research Council projects EP/P034284/1 and EP/P003990/1 (COALESCE) and the European Research Council's Advanced Fellow Grant QuantCom (Grant No. 789028)}
\thanks{Jinnan Piao, Kai Niu (\emph{Corresponding Author}), and Jincheng Dai are with the Key Laboratory of Universal Wireless Communications, Ministry of Education, Beijing University of Posts and Telecommunications (BUPT), Beijing 100876, China (email: piaojinnan@bupt.edu.cn, niukai@bupt.edu.cn, daijincheng@bupt.edu.cn).}
\thanks{Kai Niu is also with the Peng Cheng Laboratory, No.2, Xingke 1st Street, Nanshan District, Shenzhen, Guangdong Province, P. R. China.}
\thanks{Lajos Hanzo is with the School of Electronics and Computer Science, University of Southampton, Southampton SO17 1BJ, U.K. (e-mail: lh@ecs.soton.ac.uk).}
}

\maketitle

\begin{abstract}

We propose a unitary precoding scheme, namely polar-precoding, to improve the performance of polar-coded MIMO systems. In contrast to the traditional design of MIMO precoding criteria, the proposed polar-precoding scheme relies on the \emph{polarization criterion}. In particular, the precoding matrix design comprises two steps. After selecting a basic matrix for maximizing the capacity in the first step, we design a unitary matrix for maximizing the polarization effect among the data streams without degrading the capacity. Our simulation results show that the proposed polar-precoding scheme outperforms the state-of-the-art DFT precoding scheme.

\end{abstract}


\IEEEpeerreviewmaketitle

\section{Introduction}

\IEEEPARstart{T}{ransmit} precoding (TPC) is a channel-adaptive technique of precompensating the deleterious channel effects about to be encountered based on the knowledge of channel state information (CSI) at the transmitter (CSIT) \cite{MIMO}. Given the limited bandwidth of control channels in practical communication systems, typically codebook-based TPC schemes relying on a low-rate CSI-feedback are used \cite{LTE,5G}. The pivotal design aspects are the codebook design and the CSI-entry selection criterion. The simplest codebook design relies on selecting a specific antenna subset \cite{Antenna_s1,Antenna_s2}. By contrast, the Fourier codebook proposed in \cite{UP_DFT} appropriately rotates the transmit signal in a high-dimensional complex space. Furthermore, the authors of \cite{UP_STBC} and \cite{UP_SMS} transform the codebook design into packing subspaces into the Grassmann manifold relying on the projection two-norm and Fubini-Study distances, respectively. As for the CSI selection criterion, the popular capacity criterion or the maximum-likelihood (ML) criterion \cite{UP_SMS} may be used for selecting the TPC matrix from the codebook. However, these codebooks and their selection criteria were designed for uncoded MIMO systems with an emphasis on the MIMO detection performance. In reality, coded MIMO systems have to be used, where we focus on the performance of the decoded bits.

In this context, the polar-coded MIMO (PC-MIMO) systems proposed by Dai \emph{et al.} \cite{PCMIMO} have been shown to closely approach the capacity of MIMO system with the aid of successive interference cancellation (SIC), outperforming their turbo/LDPC-coded MIMO counterparts. The PC-MIMO system of \cite{PCMIMO} was designed for fast-fading channels without exploiting the CSIT. However, harnessing the knowledge of CSIT is capable of further improving the performance. Since the polarization effect of data streams is an important factor influencing the performance \cite{PCMIMO}, the PC-MIMO TPC should be designed on the basis of explicitly exploiting the polarization effect.

Table \ref{pe} boldly contrasts our novel contributions to the state-of-the-art both in terms of the selection criterion and the polarization effect, showing the novelty of this work explicitly. The polarization effect is introduced by the successive cancellation (SC) structure. The bit-polarization was first proposed by Ar{\i}kan \cite{arikan} for designing polar codes. Then, the bit-polarization was extended to symbol polarization and a $2^m$-ary multilevel polar-coded modulation scheme was proposed in \cite{Polar_coded_modulation_seidl}. Furthermore, Dai \emph{et al.} designed the PC-MIMO \cite{PCMIMO} using antenna polarization. Inspired by these papers, we conceive data stream polarization to design a unitary precoding scheme.

\begin{table*}[t]
\renewcommand\arraystretch{0.9}
\centering
\vspace{-0em}
\caption{Boldly contrasting our contributions to the state-of-the-art papers.}
\label{pe}
\begin{tabular}{|p{2.8cm}|c|c|c|c|c|c|c|c|}
\hline
&2004\cite{Antenna_s1,Antenna_s2}&2005\cite{UP_STBC}&2005\cite{UP_SMS}&2009\cite{arikan}&2013\cite{Polar_coded_modulation_seidl}&2018\cite{PCMIMO}&2020\cite{SoSCL}&This work\\
\hline
SNR criterion&\checkmark&&\checkmark&&&&&\\
\hline
ML criterion&&\checkmark&\checkmark&&&&&\\
\hline
Capacity criterion&&&\checkmark&&&&&\checkmark\\
\hline
Polarization criterion&&&&&&&&\checkmark\\
\hline
Bit polarization&&&&\checkmark&\checkmark&\checkmark&\checkmark&\checkmark\\
\hline
Symbol polarization&&&&&\checkmark&\checkmark&&\\
\hline
Antenna polarization&&&&&&\checkmark&\checkmark&\\
\hline
Data stream polarization&&&&&&&&\checkmark\\
\hline
\end{tabular}
\end{table*}

In this compact letter, a unitary polar TPC is proposed for improving the performance of PC-MIMO systems. Since the polarization of the substreams directly affects the PC-MIMO performance \cite{PCMIMO}, the proposed polar TPC scheme stems from the \emph{polarization criterion} used for maximizing the polarization effect, which constitutes a radical departure from the traditional TPC design. Given the codebook, the TPC matrix selection comprises two steps. In the first step, a basic TPC matrix is selected for maximizing the capacity. In the second step, we post-multiply the basic matrix by a unitary matrix, which is specifically designed for maximizing the polarization of substreams without eroding the capacity optimized by the basic TPC matrix.
Moreover, the optimal polar TPC of the PC-MIMO system is derived under the polarization criterion and a method to design the polar TPC codebook is proposed based on the DFT TPC.
Our simulation results illustrate that the proposed polar TPC scheme outperforms the state-of-the-art DFT TPC scheme.

\emph{Notational Conventions}: In this letter, scalars are denoted by the lowercase letters (e.g., $x$).
The calligraphic characters, such as ${\cal X}$, are used to denote sets. The bold capital letters, such as $\mathbf{X}$, denote matrices.
The $j$-th column of matrix $\mathbf{X}$ is written as ${\mathbf{X}}_j$ and $\mathbf{X}_i^j$ represents the matrix $\left[{\mathbf{X}}_i,\cdots,{\mathbf{X}}_j\right]$.
The element in the $i$-th row and the $j$-th column of matrix $\mathbf{X}$ is written as $X_{i,j}$. ${\mathbf{X}}^T$ and ${\mathbf{X}}^*$ are used to denote the transposition and the conjugate transposition of ${\mathbf{X}}$, respectively.
The bold lowercase letters (e.g., ${\bf{x}}$) are used to denote column vectors. Notation ${{x}_i^j}$ denotes the column subvector $(x_i,\cdots,x_j)^T$ and $x_i$ denotes the $i$-th element of ${\bf{x}}$.
Given an index set ${\cal A}$, $x_{\cal A}$ is a subvector composed of $x_i$, $i \in {\cal A}$.
We use ${\cal U}(M_T, M)$ to denote the set of $M_T \times M$ matrices with orthonormal columns, ${\bf I}_M$ to denote an $M \times M$ identity matrix, $\lambda_i({\bf X})$ to denote the $i$-th smallest singular value of $\bf X$ and $diag(x_1,\cdots,x_M)$ to denote an $M \times M$ diagonal matrix.
Throughout this letter, $\log \left(  \cdot  \right)$ means ``base 2 logarithm''.

\section{Preliminaries}

\subsection{Polar-Coded MIMO System}

In this section, we introduce the PC-MIMO system of \cite{PCMIMO} by intrinsically amalgamating it with a unitary TPC scheme. The $K$ information bits are first encoded and modulated into QPSK symbols, which are then precoded by the codebook and transmitted via the MIMO channel using $M_T$-transmit antennas, $M_R$-receive antennas and $M$ bitstreams within $N$ time slots. We focus on block-fading channels, where the channels remain constant for $N$ time slots.

At the transmitter, the source sequence $u_1^{2MN}$ composed of $u_{\cal A}$ and $u_{{\cal A}^c}$ with information set $|{\cal A}| = K$ and code rate $R = \frac{K}{{2MN}}$ is demultiplexed into $M$ different bitstreams and each bitstream is fed into a polar encoder. Then, the $2N$-dimensional encoded sequence $v_{1+2N(i-1)}^{2Ni}$, $1 \le i \le M$, is mapped into an $N$-dimensional modulated sequence $s_{1+N(i-1)}^{Ni}$ using QPSK modulation. Next, the $M \times N$ symbol matrix ${\bf{S}} = {[ {s_1^N,s_{N + 1}^{2N}, \cdots ,s_{N\left( {M - 1} \right) + 1}^{NM}} ]^T}$ is multiplied by a $M_T \times M$ TPC matrix $\bf F$ and produces the transmit signal matrix ${\bf X} = \sqrt {\frac{{{E_s}}}{M}} {\bf{FS}}$, where $E_s$ is the total transmit energy. Hence, the received signal matrix $\bf Y$ at the output of block fading channels is
\begin{equation}\label{symbol_matrix}
{\bf{Y}} = {\bf HX} + {\bf{Z}} = \sqrt {\frac{{{E_s}}}{M}} {\bf HF}{\bf S} + {\bf{Z}},
\end{equation}
where $\bf H$ is the channel response matrix having i.i.d entries in ${\cal CN}(0,1)$
and the elements of $\bf Z$ are i.i.d. complex circular Gaussian random variables with $z_{i,j} \sim {\cal CN}(0,N_0)$. Perfect channel estimation is assumed at the receiver.
To simplify the analysis, we rewrite the system model (\ref{symbol_matrix}) by omitting the time slot as
\begin{equation}\label{symbol_matrix_simplify}
{\bf{y}} = \sqrt {\frac{{{E_s}}}{M}} {{\bf{HF}}}{\bf{s}} + {\bf{z}}.
\end{equation}

At the receiver, a joint multistage detection and decoding receiver is used, which is similar to the SC decoding rules of polar codes \cite{arikan,Lajos1,Lajos2}. Hence, other SC-like decoding schemes, such as the successive cancellation list (SCL) decoder \cite{talvardyscl, niuscl,SoSCL}, the successive cancellation stack decoder \cite{SCS,RCSCS} and the CRC-aided SCL (CA-SCL) decoder \cite{niu_CASCL}, can also be used in the PC-MIMO system to improve the performance.
The MIMO detection order proceeds from substream $1$ to $M$. A substream is first demodulated into bit log-likelihood ratios (LLRs) and the LLRs are then fed into the polar decoder. Then, the decoded bitstream is entered into the polar encoder to retrieve the QPSK symbols. After the bits in the substream have been estimated, they are fed back to the MIMO detector in order to perform interference cancellation.

\subsection{Unitary Precoding}

The receiver selects a TPC matrix $\bf F$ from the codebook set ${\cal F}$ with $|{\cal F}| = 2^B$, where ${\cal F} \subset {\cal U}(M_T, M)$ and $B$ bits of feedback are available. The DFT-based TPC designed for spatial multiplexing systems in \cite{UP_STBC, UP_DFT} is formulated as:
\begin{equation}\label{DFTcodebook}
{\cal F} = \left\{{\bf F}_{\rm DFT}, {\bf\Theta}{\bf F}_{\rm DFT}, \cdots, {\bf\Theta}^{2^B-1}{\bf F}_{\rm DFT}\right\},
\end{equation}
where the entry of ${\bf F}_{\rm DFT}$ at $(k,l)$ is $\frac{1}{\sqrt {M_T}} e^{i\left(\frac{2\pi}{M_T} \right)kl}$ and ${\bf\Theta}$ is the diagonal matrix
 \begin{equation}\label{Theta}
{\bf{\Theta }} = diag\left({e^{i\left( {2\pi /{2^B}} \right){a_1}}}, \cdots, {e^{i\left( {2\pi /{2^B}} \right){a_{{M_T}}}}} \right).
\end{equation}
In (\ref{Theta}), the vector ${\bf a} = [a_1,\cdots,a_{M_T}]$ is:
\begin{equation}\label{vector_a}
{\bf a} = \mathop {\arg \max }\limits_{{\cal Z}} \mathop {\min }\limits_{1 \le l \le {2^B} - 1} d\left( {{{\bf{F}}_{{\rm{DFT}}}},{{\bf{\Theta }}^l}{{\bf{F}}_{{\rm{DFT}}}}} \right),
\end{equation}
where ${\cal Z} = \left\{ {\bf a}\in {\mathbb Z}^{M_T}|0 \le a_k \le 2^B-1, \forall k\right\}$ and $d({\bf A}, {\bf B}) = \frac{1}{{\sqrt 2 }}\left\| {{\bf{A}}{{\bf{A}}^*} - {\bf{B}}{{\bf{B}}^*}} \right\|$. Then, random testing of the values of ${\bf a} \in {\cal Z}$ is used to optimize the cost function for training the codebook.

The capacity maximization criterion is used to select the TPC matrix $\bf F$ from $\cal F$ yielding:
\begin{equation}\label{capacitySC}
{\bf{F}} = \mathop {\arg \max }\limits_{{\bf{F}} \in {{\cal F}}} I\left( {{\bf y};{\bf s}|{\bf{HF}}} \right),
\end{equation}
where $I\left( {{\bf y};{\bf s}|{\bf{HF}}} \right) = \log \det \left( {{{\bf{I}}_M} + \frac{{{E_s}}}{{M{N_0}}}{{\bf{F}}^*}{{\bf{H}}^*}{\bf{HF}}} \right)$ is the capacity of the unitary TPC-aided system.

\section{Polar Precoding}

In this section, we first illustrate the polarization effect of substreams. Then, the polarization criterion is provided and the optimal unquantized TPC satisfying this criterion is derived. Finally, the method of designing the polar TPC codebook is proposed.

\subsection{Polarization Effect of Substreams}

\begin{figure}[t]
\setlength{\abovecaptionskip}{0.cm}
\setlength{\belowcaptionskip}{-0.cm}
  \centering{\includegraphics[scale=1]{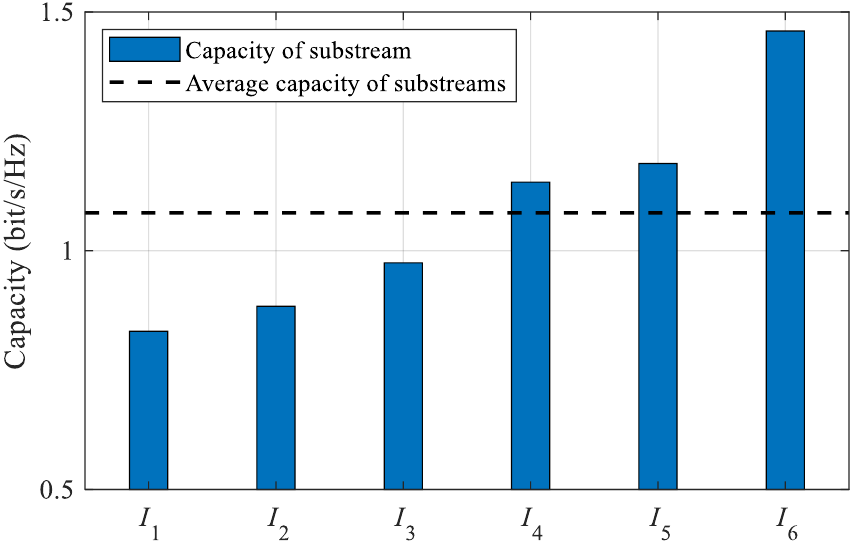}}
  \caption{The capacity of substreams relying on DFT TPC for $M_T = 8$, $M_R = 8$ and $M = 6$ at $\frac{{{E_s}}}{{{N_0}}} = 0$dB.}\label{polarization_effect}
\end{figure}

Given ${\bf G} = {\bf HF}$, the system model (\ref{symbol_matrix_simplify}) is simplified to
\begin{equation}\label{symbol_matrix_simplify_G}
{\bf{y}} = \sqrt {\frac{{{E_s}}}{M}} {{\bf{G}}}{\bf{s}} + {\bf{z}}.
\end{equation}
Then, due to the SC structure at the receiver, the system model associated with the $1$-st to the $(i-1)$-st substreams known is formulated as:
\begin{equation}\label{symbol_matrix_ith_substream}
\underbrace {{\bf{y}} - \sqrt{\frac{{{E_s}}}{M}}{\bf G}_1^{i-1}s_1^{i - 1}}_{ = :{\bf{y}}_i} = \sqrt {\frac{{{E_s}}}{M}} {{\bf G}_i^M}s_i^M + {\bf{z}}.
\end{equation}
According to the chain rule of mutual information, the capacity of the PC-MIMO system is decomposed into
\begin{equation}\label{chain_rule}
I\left( {{\bf{y}};{\bf{s}}|{\bf{HF}}} \right) = \sum\limits_{i = 1}^M {\underbrace {I\left( {{\bf{y}};{s_i}|{\bf{HF}},s_1^{i - 1}} \right)}_{ = :{I_i}}}
\end{equation}
and $I_i$ is the capacity of the $i$-th substream with SC structure, which is calculated by
\begin{equation}\label{Ii}
\begin{aligned}
{I_i} &= I\left( {{\bf{y}};s_i^M|{\bf{HF}},s_1^{i - 1}} \right) - I\left( {{\bf{y}};s_{i + 1}^M|{\bf{HF}},s_1^i} \right)\\
 &= I\left( {{\bf{y}}_i;s_i^M|{\bf{G}}_i^M} \right) - I\left( {{\bf{y}}_{i + 1} ;s_{i + 1}^M|{\bf{G}}_{i + 1}^M} \right).
\end{aligned}
\end{equation}

Similar to \cite{arikan, PCMIMO}, the SC structure also introduces the polarization effect of substreams, i.e., the capacity difference among $I_i, i=1,\cdots, M$. Fig. \ref{polarization_effect} is an example illustrating the polarization effect for $M_T = 8$, $M_R = 8$ and $M = 6$ at $\frac{{{E_s}}}{{{N_0}}} = 0$dB using DFT TPC. In Fig. \ref{polarization_effect}, the capacities of the $M = 6$ substreams are increasing from $I_1$ to $I_6$.
Based on that, $I_1$, $I_2$ and $I_3$ are lower than the average capacity and $I_4$, $I_5$ and $I_6$ are higher than the average capacity.
Thus, a capacity difference occurs among $I_1$ to $I_6$ and the polarization effect is introduced by the SC structure.

\subsection{Polarization Criterion}

In PC-MIMO systems, drastic polarization leads to better performance when the capacity is identical \cite{PCMIMO}.
Thus, maximizing the system capacity
\begin{equation}\label{max_cap}
{\bf F} = \mathop {\arg \max }\limits_{{\bf F} \in {\cal F}} I\left( {\bf{y};\bf{s}|{\bf{HF}}} \right)
\end{equation}
and simultaneously maximizing the polarization effect among the substreams
\begin{equation}\label{max_polar}
{\bf{F}} = \mathop {\arg \max }\limits_{{\bf{F}} \in {{\cal F}}}  {\sum\limits_{i = 1}^M {{{\left( {{I_i} - \bar I} \right)}^2}} }
\end{equation}
are both necessary for our polar TPC, where $\bar I$ is the average capacity of the $M$ substreams, i.e., $\bar I = \frac{{I\left( {{\bf{y}};{\bf{s}}|{\bf{HF}}} \right)}}{M}$.
However, it is a challenge to directly find a suitable $\bf F$ satisfying both (\ref{max_cap}) and (\ref{max_polar}).

Then, since $I\left( {\bf{y};\bf{s}|{\bf{HF}}} \right)$ remains unchanged when $\bf F$ is multiplied by a unitary matrix, $\bf F$ is partitioned into two matrices ${\bf W} \in {\cal U}(M_T,M)$ and ${\bf Q} \in {\cal U}(M,M)$, and we have
\begin{equation}\label{cap_unchanged}
I\left( {\bf{y};\bf{s}|{\bf{HF}}} \right) = I\left( {\bf{y};\bf{s}|{\bf{HWQ}}} \right) = I\left( {\bf{y};\bf{s}|{\bf{HW}}} \right),
\end{equation}
where ${\bf F} = {\bf WQ}$.

Based on \eqref{cap_unchanged}, we can find a matrix ${\bf W}$ for maximizing $I\left( {\bf{y};\bf{s}|{\bf{HW}}} \right)$, which is equivalent to maximizing $I\left( {\bf{y};\bf{s}|{\bf{HF}}} \right)$.
When $\bf W$ is determined, $I\left( {\bf{y};\bf{s}|{\bf{HF}}} \right)$ remains unchanged for $\forall{\bf Q} \in {\cal U}(M,M)$. Thus, $\bf Q$ can be used for maximizing the polarization effect without affecting the system capacity.
Hence, ${\bf F} = {\bf WQ}$ can satisfy both (\ref{max_cap}) and (\ref{max_polar}). The polarization criterion is defined as
\begin{equation}\label{criterion_new}
\left\{
\begin{aligned}
\bf{W} &= \mathop {\arg \max }\limits_{{\bf{W}} \in {\cal W}} I\left( {\bf{y};\bf{s}|{\bf{HW}}} \right)\\
{\bf{Q}} &= \mathop {\arg \max }\limits_{{\bf{Q}} \in {{\cal Q}}} {\sum\limits_{i = 1}^M {{{\left( {{I_i} - \bar I} \right)}^2}} },
\end{aligned}
\right.
\end{equation}
where ${\cal W} \subset {\cal U}(M_T,M)$ and ${\cal Q} \subset {\cal U}(M,M)$ are the codebooks for $\bf W$ and $\bf Q$, respectively.

\subsection{Optimal Unquantized TPC}

According to the polarization criterion, the system model (\ref{symbol_matrix_simplify}) is transformed into
\begin{equation}\label{symbol_matrix_criterion}
{\bf{y}} = \sqrt {\frac{{{E_s}}}{M}} {\bf HWQ}{\bf s} + {\bf{z}}.
\end{equation}
Let the singular value decomposition of a matrix $\bf A$ be given by
\begin{equation}\label{SVD}
{\bf A} = {\bf U}_{\bf A}{\bf \Sigma}_{\bf A}{\bf V}_{\bf A}^*,
\end{equation}
where ${\bf U}_{\bf A}$ and ${\bf V}_{\bf A}$ are unitary matrices and ${\bf \Sigma}_{\bf A}$ is a diagonal matrix with $\lambda_k({\bf A})$ denoting the $k$-th smallest singular value of $\bf A$ at entry $(k,k)$.

Then, based on (\ref{symbol_matrix_criterion}), we first derive ${\bf Q}_{opt} \in {\cal U}(M,M)$ that maximizes the polarization effect with $\bf W$.

\begin{lemma}\label{lemma1}
The optimal TPC matrix ${\bf Q}_{opt} \in {\cal U}(M,M)$ with $\bf W$ is ${\bf Q}_{opt} = {\bf V}_{\bf HW}$.
\end{lemma}
\begin{IEEEproof}
For the system model (\ref{symbol_matrix_ith_substream}), we have ${\bf G}_i^M = {\bf HW}{\bf Q}_i^M$. In \cite{UP_SMS}, it has been proved that ${\bf Q}_i^M = {{\bf V}_{\bf HW}}_i^M$ can maximize $I\left( {{\bf{y}}_i;s_i^M|{\bf{G}}_i^M} \right)$, where ${{\bf V}_{\bf HW}}_i^M$ is a matrix constructed from the last $(M-i+1)$ columns of ${\bf V}_{\bf HW}$.
Thus, ${\bf Q}_{opt}$ maximizes $I\left( {{\bf{y}}_i;s_i^M|{\bf{G}}_i^M} \right) =
\sum\nolimits_{k = i}^M {{I_k}}, i=1,\cdots,M$.

Let $I_k$ denote the capacity of the $k$-th substream optimized by ${\bf Q}_{opt}$ and $I_1 \le I_2 \le \cdots \le I_M$.
We transform the proof into linear programming as follows:
\begin{equation}\label{LP}
\begin{aligned}
{\max}~& f\left(x_1,x_2, \cdots, x_M\right) = {\sum\limits_{k = 1}^M {{{\left( {{x_k} - \bar I} \right)}^2}} },\\
{\rm{s. t.}~}& {\sum\limits_{k = i}^M {x_k} } \le \sum\limits_{k = i}^M {{I_k}}, 2 \le i \le M,\\
&{\sum\limits_{k = 1}^M {x_k} } = M\bar I.
\end{aligned}
\end{equation}
Then, since $f\left(x_1,x_2, \cdots, x_M\right)$ is a convex function, the maximum value is on the boundary and the point is $x_k = I_k$, $1 \le k \le M$.
Thus, ${\bf Q}_{opt} = {\bf V}_{\bf HW}$ is the optimal TPC matrix with $\bf W$.
\end{IEEEproof}

According to Lemma \ref{lemma1}, we can readily derive the optimal TPC matrix ${\bf F}_{opt}$ for satisfying the polarization criterion.
\begin{lemma}\label{lemma2}
The optimal TPC matrix ${\bf F}_{opt} \in {\cal U}(M_T,M)$ is a matrix constructed from the last $M$ columns of ${\bf V}_{\bf H}$.
\end{lemma}
\begin{IEEEproof}
In \cite{UP_SMS}, it has been shown that ${\bf F}_{opt}{\bf Q}$ is the optimal TPC maximizing $I\left( {{\bf{y};\bf{s}}|{{\bf HF}_{opt}{\bf Q}}} \right)$, where ${\bf F}_{opt}$ is composed of the last $M$ columns of ${\bf V}_{\bf H}$ and $\forall {\bf Q} \in {\cal U}(M,M)$. Then, according to Lemma \ref{lemma1}, the optimal polar TPC associated with fixed ${\bf F}_{opt}$ is
${\bf Q}_{opt} = {\bf V}_{{\bf HF}_{opt}} = {\bf I}_M$. Thus, the optimal polar TPC matrix is ${\bf F}_{opt}$, which is constructed from the last $M$ columns of ${\bf V}_{\bf H}$.
\end{IEEEproof}

\subsection{Polar TPC Codebook Design}

The codebook of the polar TPC is ${\cal F} = \left\{{\bf F}|{\bf F = WQ},{\bf W}\in{\cal W}, {\bf Q}\in{\cal Q}\right\}$ with $|{\cal F}| = 2^{B}$, $|{\cal W}| = 2^{B_1}$, $|{\cal Q}| = 2^{B_2}$ and $B = B_1 + B_2$. The codebook design is divided into two steps, which are summarized as follows:
\begin{enumerate}
\item $\cal W$ is designed by the DFT TPC of \cite{UP_STBC, UP_DFT}, i.e.,
\begin{equation}\label{WDFTcodebook}
{\cal W} = \left\{{\bf W}_{\rm DFT}, {\bf\Theta}_{\bf W}{\bf W}_{\rm DFT}, \cdots, {\bf\Theta}_{\bf W}^{2^{B_1}-1}{\bf W}_{\rm DFT}\right\},
\end{equation}
where the entry of ${\bf W}_{\rm DFT}$ at $(k,l)$ is
$\frac{1}{\sqrt {M_T}} e^{i\left(\frac{2\pi}{M_T} \right)kl}$ and the diagonal matrix ${\bf\Theta}_{\bf W}$ is
 \begin{equation}\label{WDFT_Theta}
{\bf{\Theta }_{\bf W}} = diag\left({e^{i\left( {2\pi /{2^{B_1}}} \right){a_1}}}, \cdots, {e^{i\left( {2\pi /{2^{B_1}}} \right){a_{{M_T}}}}} \right).
\end{equation}
The vector ${\bf a} = [a_1,\cdots,a_{M_T}]$ in (\ref{WDFT_Theta}) is
\begin{equation}\label{vector_a_W}
{\bf a} = \mathop {\arg \max }\limits_{{\cal Z}} \mathop {\min }\limits_{1 \le l \le {2^{B_1}} - 1} d\left( {{{\bf{W}}_{{\rm{DFT}}}},{{\bf{\Theta }}_{\bf W}^l}{{\bf{W}}_{{\rm{DFT}}}}} \right).
\end{equation}

\item $\cal Q$ is also designed by the DFT TPC, i.e.,
 \begin{equation}\label{QDFTcodebook}
{\cal Q} = \left\{{\bf Q}_{\rm DFT}, {\bf\Theta}_{\bf Q}{\bf Q}_{\rm DFT}, \cdots, {\bf\Theta}_{\bf Q}^{2^{B_2}-1}{\bf Q}_{\rm DFT}\right\}.
\end{equation}
Hence, the entry of ${\bf Q}_{\rm DFT}$ at $(k,l)$ is $\frac{1}{\sqrt {M}} e^{i\left(\frac{2\pi}{M} \right)kl}$ and the diagonal matrix ${\bf\Theta}_{\bf Q}$ is
  \begin{equation}\label{QDFT_Theta}
{\bf{\Theta }_{\bf Q}} = diag\left(1,{e^{i\left( {2\pi /{2^{B_2}}} \right)}}, \cdots, {e^{i\left( {2\pi /{2^{B_2}}} \right){\left(M-1\right)}}} \right).
\end{equation}

\end{enumerate}

For polar TPC, the optimization of $B_1$ and $B_2$ is important. In this paper, $B_1$ and $B_2$ are selected empirically and we just provide a compact insight into the optimization.
According to the polarization criterion (\ref{criterion_new}), $B_1$ and $B_2$ affect the capacity and the polarization effect, respectively.
Then, a higher $B_1$ or a lower $B_2$ leads to higher capacity and lighter polarization effect, and vice versa.
Explicitly, both factors have an influence on the PC-MIMO performance. Thus, the polar TPC codebook has to strike a trade-off between the capacity and the polarization effect, and both $B_1$ as well as $B_2$ should be optimized.

\section{Performance Evaluation}

\begin{figure}[t]
\setlength{\abovecaptionskip}{0.cm}
\setlength{\belowcaptionskip}{-0.cm}
  \centering{\includegraphics[scale=1]{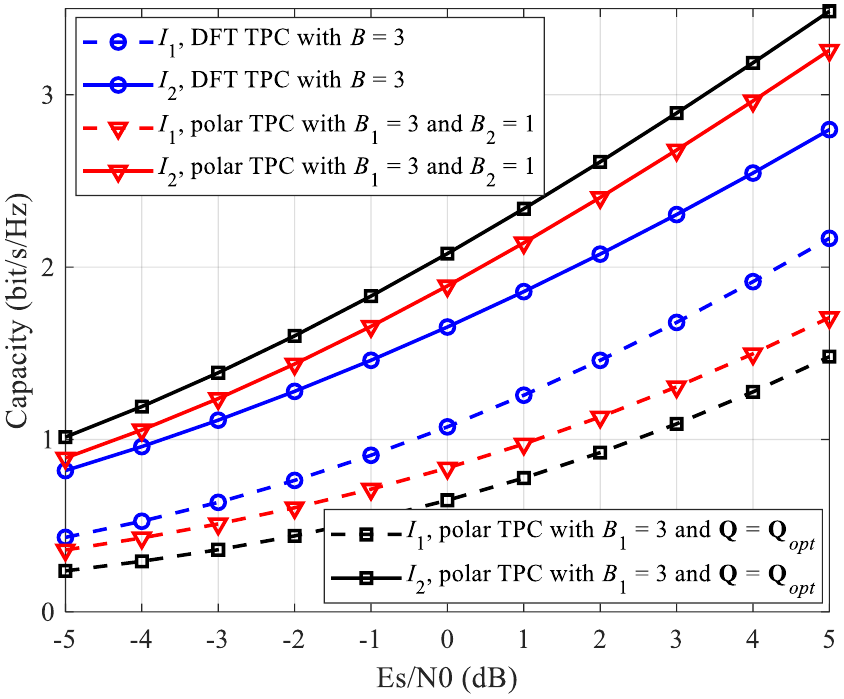}}
  \caption{The capacity of substreams for the fixed channel response of (\ref{H}) and different TPCs, where $M_T = 3$, $M_R = 3$ and $M = 2$.}\label{capacity}
\end{figure}

\begin{figure}[t]
\setlength{\abovecaptionskip}{0.cm}
\setlength{\belowcaptionskip}{-0.cm}
  \centering{\includegraphics[scale=1]{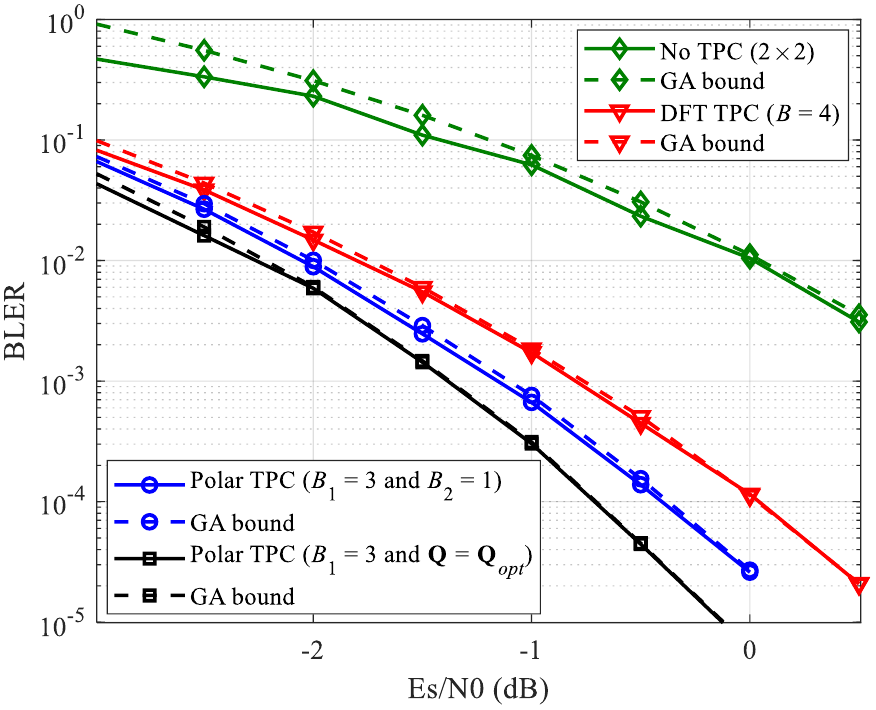}}
  \caption{The BLER of PC-MIMO systems using different TPC schemes, where $M_T = 3$, $M_R = 3$, $M = 2$, $N = 64$ and $R = 1/4$.}\label{BLER1}
\end{figure}

In this section, we first provide the capacity of the substreams for the fixed channel matrix
\begin{equation}\label{H}
{\bf{H}} = \left[ {\begin{array}{*{20}{c}}
{0.61 - 0.92i}&{-0.93 + 0.56i}&{-1.24 + 0.35i}\\
{0.93 - 1.30i}&{-0.21 - 0.15i}&{-0.51 - 0.60i}\\
{0.01 + 0.35i}&{-0.64 - 0.44i}&{0.78 + 0.04i}
\end{array}} \right].
\end{equation}
Then, the block error rate (BLER) performance of the proposed polar TPC is provided for the channel response in (\ref{H}). Finally, we provide the BLER performance of polar TPC under block-fading channels. The PC-MIMO system is constructed by the Gaussian approximation (GA) \cite{GA_Trifonov}.
%Then, to match the polarization criterion, ML-SIC detection is used for the PC-MIMO system.
The polarization criterion maximizes both the capacity and the polarization effect simultaneously.
To allow the system performance approach the capacity, ML detection is considered.
Furthermore, since the polarization effect is catalyzed by the SIC structure of the PC-MIMO system, ML-SIC detection is used in this paper.

Fig. \ref{capacity} shows the capacity of the substreams for different TPC schemes and for the fixed channel response in (\ref{H}), where $M_T = 3$, $M_R = 3$ and $M = 2$. In Fig. \ref{capacity}, $I_1$ and $I_2$ are the capacities of the first and the second substreams, respectively. We can observe that by introducing the TPC matrix $\bf Q$, the polarization effect of the polar TPC for $B_1 = 3$ and $B_2 = 1$ is higher than that of the DFT TPC with $B = 3$. Thus, the proposed polar TPC enhances the polarization effect among the substreams, which improves the BLER of the PC-MIMO system shown in Fig. \ref{BLER1}.
Then, the polar TPC using the optimal TPC ${\bf Q}_{opt}$ also shows more significant polarization effect compared to the polar TPC with $B_1 = 3$ and $B_2 = 1$. Similarly, the more significant polarization effect improves the BLER in Fig. \ref{BLER1} as well.

Fig. \ref{BLER1} illustrates the BLER of PC-MIMO systems for different TPC schemes, where $M_T = 3$, $M_R = 3$, $M = 2$, $N = 64$ and $R = 1/4$.
ML-SIC detection and SC decoding are used for the PC-MIMO system.
Then, in order to make the comparison fair, the performance of the DFT and polar TPCs having identical number of  feedback bits is provided, i.e., $B = 4$ for the DFT TPC, and $B_1 = 3$ as well as $B_2 = 1$ for the polar TPC. In Fig. \ref{BLER1}, we can first observe that the GA bound, widely used in \cite{PCMIMO, GA_Trifonov}, is still an upper bound of the performance of PC-MIMO TPC schemes under SC decoding. Moreover, the GA bound coincides with the corresponding BLER performance in the high signal-to-noise ratio (SNR) regions.
Furthermore, as expected, both the DFT and the polar TPCs outperform the ``no-TPC'' system. Hence, TPC efficiently improves the performance of PC-MIMO. Additionally, since the proposed polar TPC has better polarization effect than the DFT TPC, it has about $0.45$dB performance gain at BLER $10^{-4}$.
Moreover, due to the better polarization effect shown in Fig. \ref{capacity}, the performance of the polar TPC relying on the optimal TPC ${\bf Q}_{opt}$ achieves about $0.4$dB gain over the polar TPC with $B_1 = 3$ and $B_2 = 1$ at BLER $10^{-4}$. Therefore, better polarization leads to a better PC-MIMO performance using the proposed polar TPC instead of other known TPCs.

\begin{figure}[t]
\setlength{\abovecaptionskip}{0.cm}
\setlength{\belowcaptionskip}{-0.cm}
  \centering{\includegraphics[scale=1]{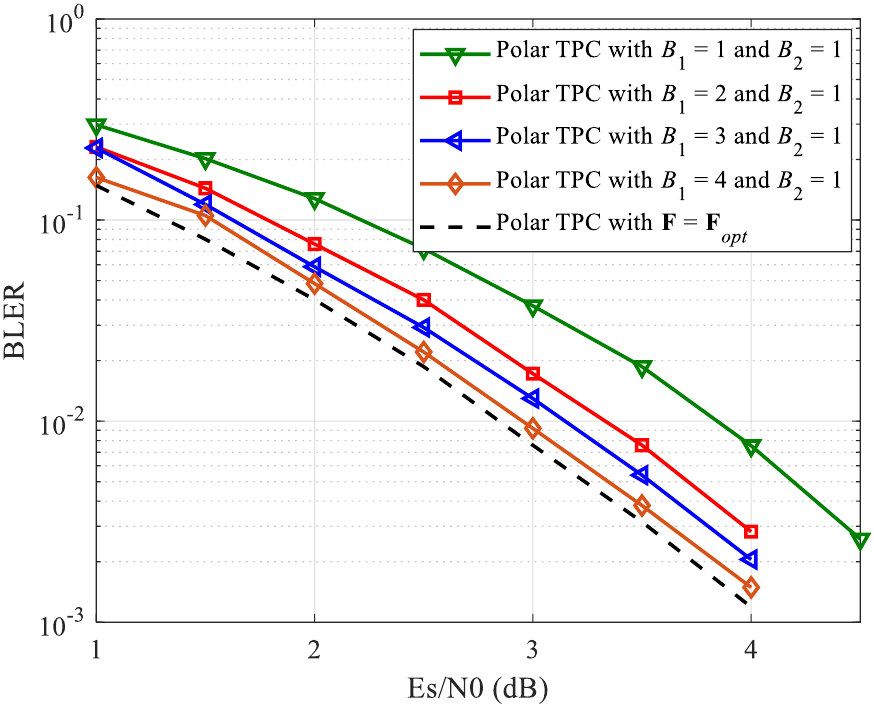}}
  \caption{The BLER of PC-MIMO systems using CA-SCL decoder and polar TPC, where $M_T = 4$, $M_R = 4$, $M = 3$, $N = 128$, $R = 1/2$ and the list size of CA-SCL is 8.}\label{BLER2}
\end{figure}

Fig. \ref{BLER2} provides the BLER of PC-MIMO systems using CA-SCL decoding \cite{niu_CASCL} and polar TPC under block-fading channels, where $M_T = 4$, $M_R = 4$, $M = 3$, $N = 128$ and $R = 1/2$. The ML-SIC MIMO detection is used and the list size of the CA-SCL decoder is 8, where the 6-bit CRC of \cite{3GPP_5G_polar} is used. In Fig. \ref{BLER2}, the performance of the polar TPC using the optimal TPC ${\bf F}_{opt}$ is provided,
which can be treated as the best-case bound of the polar TPC, since ${\bf F}_{opt}$ maximizes the polarization effect of polar TPC. Then, we can observe that the performance of polar TPC using limited feedback is close to the performance of polar TPC using ${\bf F}_{opt}$ as $B_1$ increases.
Specifically, the polar TPC using $B_1 = 4$ and $B_2 = 1$ has almost identical BLER to that of ${\bf F}_{opt}$ in the high SNR regions. Thus, the polar TPC has the potential of approaching the optimal performance, despite of limited feedback.

\begin{figure}[t]
\setlength{\abovecaptionskip}{0.cm}
\setlength{\belowcaptionskip}{-0.cm}
  \centering{\includegraphics[scale=0.67]{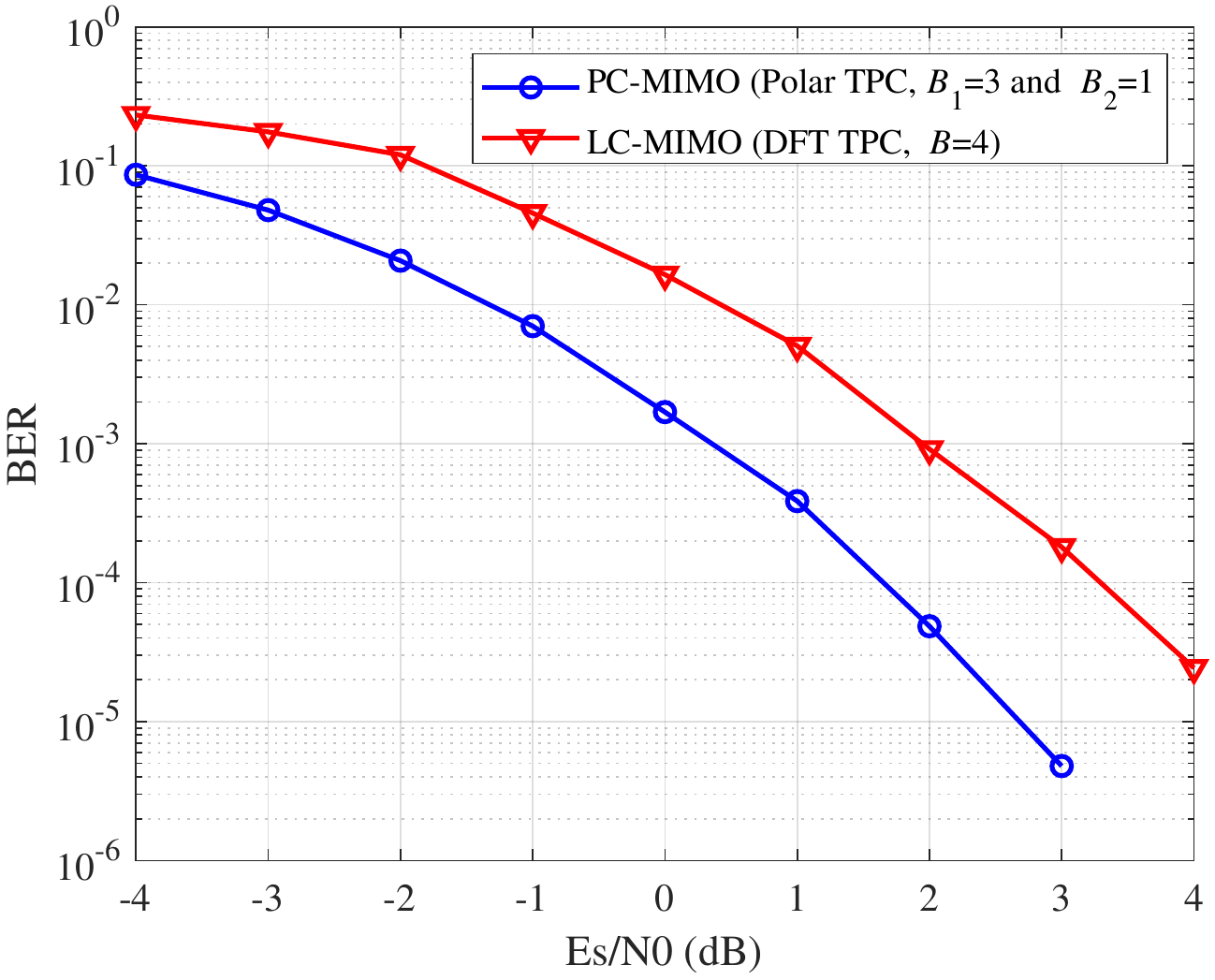}}
  \caption{The BER comparison between PC-MIMO system with polar TPC and LC-MIMO system with DFT TPC, where $M_T = 4$, $M_R = 4$, $M = 3$, $N = 64$ and $R = 1/3$. }\label{BLE_LDPC}
\end{figure}

\begin{figure}[t]
\setlength{\abovecaptionskip}{0.cm}
\setlength{\belowcaptionskip}{-0.cm}
  \centering{\includegraphics[scale=0.67]{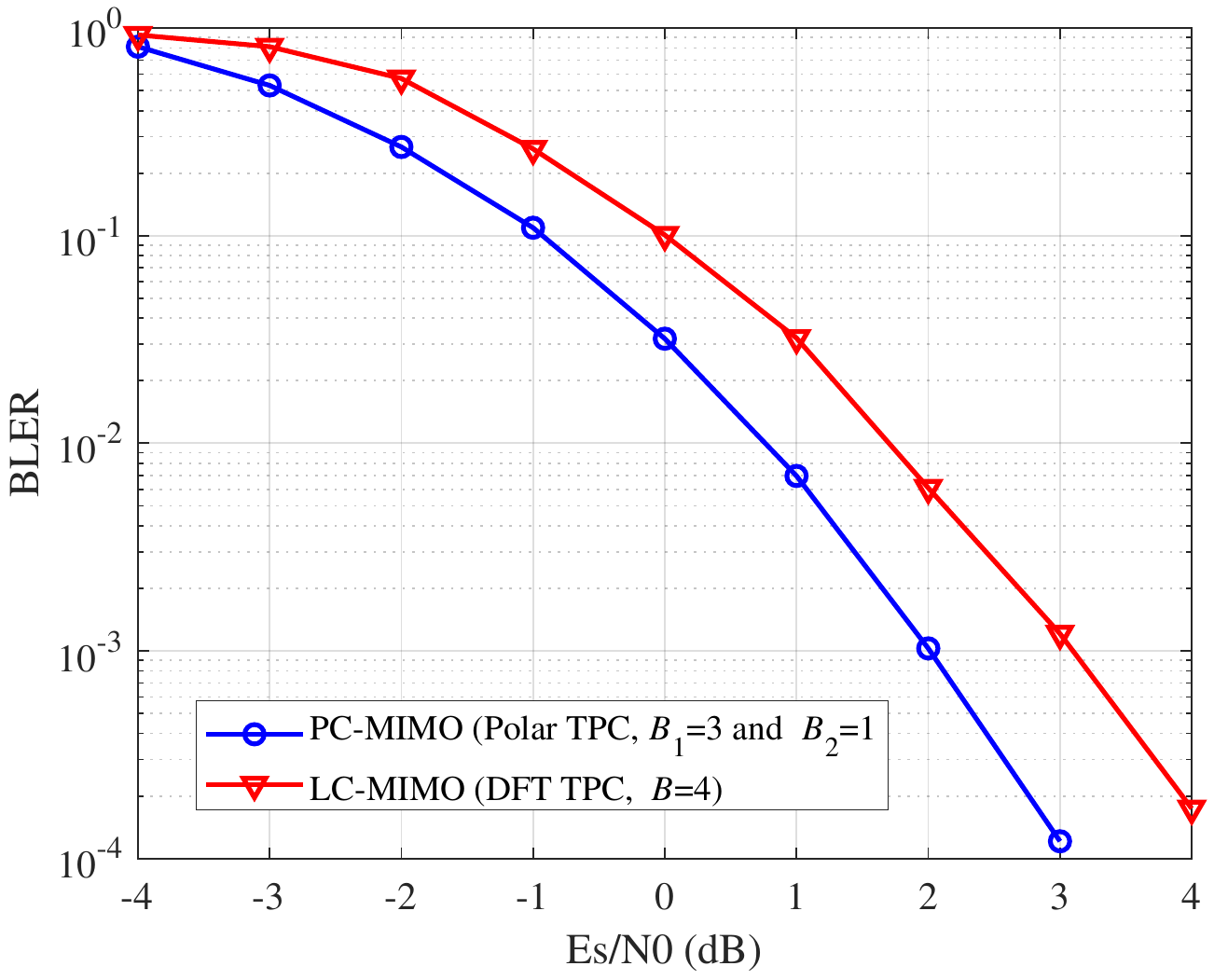}}
  \caption{The BLER comparison between PC-MIMO system with polar TPC and LC-MIMO system with DFT TPC, where $M_T = 4$, $M_R = 4$, $M = 3$, $N = 64$ and $R = 1/3$. }\label{BLER_LDPC}
\end{figure}

Fig. \ref{BLE_LDPC} and Fig. \ref{BLER_LDPC} portray out BER and BLER performance comparisons, respectively, where we have $M_T = 4$, $M_R = 4$, $M = 3$, $N = 64$ and $R = 1/3$.
For the PC-MIMO system, the CA-SCL decoder having a list size of 8 and 6-bit CRC \cite{3GPP_5G_polar} is used, where the MIMO detector is ML-SIC. For the low-density-parity-check (LDPC)-coded MIMO (LC-MIMO) system, the LDPC encoder and the rate-matching algorithm are those of 5G \cite{3GPP_5G_polar}, the sum-product algorithm having 25 iterations and layered scheduling are used for the LDPC decoder \cite{LinShu}, and the MIMO detector uses the linear minimum mean square error (LMMSE) algorithm.
In Fig. \ref{BLE_LDPC} and Fig. \ref{BLER_LDPC}, we can observe that the PC-MIMO system using polar TPC has better BER and BLER performance than the LC-MIMO system associated with DFT TPC. Specifically, at BER $10^{-4}$ and BLER $10^{-3}$, the PC-MIMO system has 1.6dB and 1.1dB performance gain over the LC-MIMO system, respectively.

Since the codebook is designed offline, selecting an appropriate precoding matrix from the codebook dominates the complexity of precoding.
The complexity of calculating the capacity is on the order of $O\left(M_TM_RM\right)$.
Hence, the complexity of the DFT TPC relying on the capacity criterion is $O\left(2^BM_TM_RM\right)$. For polar TPC, the complexities of selecting $\bf W$ and $\bf Q$ are $O\left(2^{B_1}M_TM_RM\right)$ and
$O\left(2^{B_2}M_TM_RM^2\right)$, respectively. Then, the complexity of polar TPC is $O\left((2^{B_1} + 2^{B_2}M )M_TM_RM\right)$.

\section{Conclusion}

In this compact letter, we proposed the polar TPC of PC-MIMO systems relying on the new polarization criterion, which is quite different from other design criteria. Based on this new polarization criterion, the optimal TPC was derived and the method of designing the polar TPC codebook was proposed. The simulation results illustrate that the proposed polar TPC outperforms its DFT-based counterpart.

\end{document}